\documentclass[12pt]{article}
\usepackage{amsfonts}
\usepackage{amssymb,amsmath,amsthm,latexsym}
\usepackage[numbers,compress]{natbib}
\usepackage{amsmath}
\usepackage{indentfirst}
\allowdisplaybreaks[4]
\newtheorem{theorem}{Theorem}[section]

\newtheorem{lemma}[theorem]{Lemma}

\newtheorem{define}[theorem]{Definition}
\newtheorem{example}[theorem]{Example}

\newcommand{\EOP} { \hfill $\Box$ }

\newcommand{\pf} { {\rm \noindent{\bf Proof.}} }

\numberwithin{equation}{section}

\begin{document}
\title{  A class of $p$-ary cyclic codes and their weight enumerators }
\author{Long Yu, Hongwei Liu{\thanks{Corresponding author. \newline  Email addresses:~hwliu@mail.ccnu.edu.cn~(Hongwei Liu),~longyu@mails.ccnu.edu.cn~(Long Yu).} }}
\date{School of Mathematics and Statistics, Central China Normal University, Wuhan, Hubei 430079, China}
\maketitle
\begin{abstract}
Let $m$, $k$ be  positive integers such that $\frac{m}{\gcd(m,k)}\geq 3$, $p$ be an odd prime and $\pi $ be a primitive element
of $\mathbb{F}_{p^m}$. Let $h_1(x)$ and $h_2(x)$ be the minimal polynomials of $-\pi^{-1}$ and $\pi^{-\frac{p^k+1}{2}}$ over $\mathbb{F}_p$, respectively. In the case of odd $\frac{m}{\gcd(m,k)}$, when  $k$ is even, $\gcd(m,k)$ is odd or when  $\frac{k}{\gcd(m,k)}$ is odd, Zhou et~al. in \cite{zhou} obtained the weight distribution of a class of cyclic codes $\mathcal{C}$ over $\mathbb{F}_p$ with parity-check polynomial $h_1(x)h_2(x)$. In this paper, we further investigate  this class of  cyclic codes $\mathcal{C}$ over $\mathbb{F}_p$     in the rest case of odd $\frac{m}{\gcd(m,k)}$ and the case of even $\frac{m}{\gcd(m,k)}$. Moreover, we determine the weight distribution of cyclic codes $\mathcal{C}$.
\end{abstract}



\section{Introduction}
Let $p$ be an odd prime and $q$ be a power of $p$. An $[n,k,d]$ linear code over the finite field $\mathbb{F}_q$ is a $k$-dimensional subspace of $\mathbb{F}_{q}^{n}$ with minimum Hamming distance $d$.  Let $A_i$ denote the number of codewords with Hamming weight $i$ in a linear code $\mathcal{C}$ of length $n$. The weight enumerator of $\mathcal{C}$ is defined by
\[A_0+A_1X+A_2X^2+\cdots+A_{n}X^{n}, \ \ \ \ {\rm where}\ A_0=1.\]
The sequence $(A_0,A_1,\cdots,A_n)$ is called the weight distribution of the code $\mathcal{C}$.

A linear code $\mathcal{C}$ of length $n$ is called cyclic if $(c_0,c_1,\cdots,c_{n-1})\in \mathcal{C}$ implies  $ (c_{n-1},c_0,\cdots,c_{n-2}) \in \mathcal{C}$. By identifying a codeword $(c_0,c_1,\cdots,c_{n-1})\in \mathcal{C}$ with the polynomial
\[c_0+c_1X+\cdots+c_{n-1}X^{n-1}\in \mathbb{F}_q[X]/(X^n-1),\]
a cyclic code $\mathcal{C}$ of length $n$ over $\mathbb{F}_q$ corresponds to an ideal of $\mathbb{F}_q[X]/(X^n-1)$. The monic generator $g(X)$ of this ideal is called the generator polynomial of $\mathcal{C}$, which satisfies that $g(X)|(X^n-1)$. The polynomial $h(X)=(X^n-1)/g(X)$ is referred to as the parity-check polynomial of $\mathcal{C}$ \cite{macwilliams}. In general the weight
distribution of cyclic codes are difficult to be determined and they are known only for a few special classes. There are some results on  the weight distribution of
cyclic codes whose duals have two or more zeros (see \cite{Ding2011,Ding2013,FengandLuo,Feng,LuoandFeng1,LuoandFeng2,Rao,Vega12,Vega07,Wang,XiongDCC,Xiong2013,Xiong2012,Longyu,Zeng2012,Zeng,zheng,zhou,zhouIT}, and the references therein).

The following notations are fixed throughout this paper.
\begin{itemize}
  \item Let $p$ be an odd prime, $m$, $k$ be  positive integers, $d=\gcd(m,k)$, $s=\frac{m}{d}\geq 3$, $q=p^d$ and $q^*=(-1)^{\frac{q-1}{2}}q$.
  \item Let $\mathbb{F}_{p^m}$ be the finite field and $\mathbb{F}_{p^m}^*=\mathbb{F}_q \setminus \{0\}$, $\pi$ be a primitive element of $\mathbb{F}_{p^m}$. For a given divisor $l$ of $m$, the
trace function from $\mathbb{F}_{p^m}$ to $\mathbb{F}_{p^l}$ is defined by ${\rm Tr}_l^m(x)=\sum_{i=0}^{\frac{m}{l}-1}x^{p^{li}}$.
\item Let  $v_2(j)$ denote the  $2$-adic valuation of integer $j$ (i.e., the maximal power of $2$ dividing $j$).
  \item Let $SQ$ denote the set of  square elements in $\mathbb{F}_{p^m}^*$,  $SQ_p$ denote the set of square elements in $\mathbb{F}_{p}^*$,   $u_p$ be a primitive element in $\mathbb{F}_p$, and $\zeta_p$ be a primitive  $p$-th unity root.
\end{itemize}

Let $h_1(x)$ and $h_2(x)$ be the minimal polynomials of $-\pi^{-1}$ and $\pi^{-\frac{p^k+1}{2}}$ over $\mathbb{F}_p$, respectively. It is easy to check that $h_1(x)$ and $h_2(x)$ are polynomials of degree $m$ and are pairwise distinct.
In this paper, we let  $\mathcal{C}$
be a cyclic code with parity-check polynomial $h_1(x)h_2(x)$, and then $\dim_{\mathbb{F}_p}\mathcal{C}=2m$.
By the well-known Delsarte's Theorem~\cite{Delsarte}, cyclic code $\mathcal{C}$ can be expressed as
\begin{equation}\label{eq:cycliccode}
\mathcal{C}=\left\{\textbf{c}(\alpha,\beta)=\left({\rm Tr}_1^m(\alpha\pi^{\frac{p^k+1}{2}i}+\beta(-\pi)^{i})\right)_{i=0}^{p^m-2}\mid \alpha,\beta \in \mathbb{F}_{p^m}\right\}.
\end{equation}
Recently,   Zhou et~al. in \cite{zhou}  studied this class of cyclic codes  $ \mathcal{C}$ in the case of odd $s$, when  $k$ is even, $d$ is odd or when  $\frac{k}{d}$ is odd. They showed    cyclic codes $ \mathcal{C}$   have only three nonzero weights. In this paper, we further investigate this class of cyclic codes $ \mathcal{C}$ for the rest cases, and determine the weight distribution of $ \mathcal{C}$.

The aim of this paper is to determine the weight distribution of a class of cyclic codes $\mathcal{C}$ over $\mathbb{F}_p$ defined by (\ref{eq:cycliccode}).  To this end, by using the  value distribution of the exponential sum $\sum\limits_{x\in \mathbb{F}_{p^m}}\zeta_p^{{\rm Tr}_1^m(\alpha x^{p^k+1}+\beta x^2)},\ \alpha,\beta \in \mathbb{F}_{p^m}$ (see {Theorem~$1$}\cite{LuoandFeng2}) and investigating some overdeterminate equations over finite fields, we obtain the value distribution of the exponential sum $\sum\limits_{x\in \mathbb{F}_{p^m}}(\zeta_p^{{\rm Tr}_1^m(\alpha x^{p^k+1}+\beta x^2)}+\zeta_p^{{\rm Tr}_1^m(\alpha\pi^{\frac{p^k+1}{2}} x^{p^k+1}-\beta \pi x^2)}), \ \alpha,\beta \in\mathbb{F}_{p^m}$. Applying these results,  the weight distribution of  cyclic codes $\mathcal{C}$ over $\mathbb{F}_p$ defined by (\ref{eq:cycliccode}) is obtained.

This paper is organized as follows.  Section $\textbf{2}$  gives some preliminary results. In Section $\textbf{3}$,  we give the weight distribution of a class of cyclic codes $\mathcal{C}$ defined by~(\ref{eq:cycliccode})  over $\mathbb{F}_p$.

\section{Preliminaries }
In the following, we give a brief introduction to the theory of quadratic forms over finite fields, which is needed to calculate the weight distribution of  cyclic codes $\mathcal{C}$ in
the next section.
\begin{define}
Let $\{\omega_1, \omega_2, \cdots, \omega_s\}$ be a basis for $\mathbb{F}_{q^s}$
over $\mathbb{F}_q$ and $x=\sum_{i=1}^s x_i\omega_i$, where $x_i \in \mathbb{F}_q$. A function $f(x)$ from $\mathbb{F}_{q^s}$
to $\mathbb{F}_q$ is called a quadratic form if it can be
represented as $$f(x)=f(\sum_{i=1}^s x_i\omega_i)=\sum_{1\leq i\leq j\leq s}a_{ij}x_ix_j ,\ \ \ \ a_{ij}\in \mathbb{F}_q.$$
\end{define}
\noindent The rank of the quadratic form $f(x)$ is defined as the codimension of
the $\mathbb{F}_q$-vector space
\[ V=\{ x\in \mathbb{F}_{q^s} \mid f(x+z)-f(x)-f(z)=0,\ \ {\rm for \ \ all }\ z\in \mathbb{F}_{q^s}\}, \] denotes by rank $r$. Then $|V|=q^{s-r}$.

For a quadratic form $f (x)$ with $s$ variables over $\mathbb{F}_q$, there exists a symmetric matrix $A$ of order $s$ over $\mathbb{F}_q$ such that $f(x)=XAX'$, where $X=(x_1,x_2,\cdots,x_s)\in \mathbb{F}_q^s$ and $X'$
denotes the transpose of $X$. It is known that there exists a nonsingular matrix $T$ over $\mathbb{F}_q$ such that $TAT'$ is a diagonal matrix \cite{Lidl R}. Making a nonsingular linear substitution $X=ZT$ with $Z=(z_1,z_2,\cdots,z_s)\in \mathbb{F}_q^s$, we have
\begin{equation}\label{eq:f(x)=Z(TAT')Z'}
f(x)=Z(TAT')Z'=\sum_{i=1}^ra_iz_i^2,\ \ \ \ a_i\in \mathbb{F}_q^*,
\end{equation}
where $r$ is the rank of $f(x)$. We have the following result (see Lemma~$1$ \cite{LuoandFeng2}).
\begin{lemma}\label{lem:erciquzhi}{\rm \cite{LuoandFeng2}}
Let $\eta_d$ be the quadratic (multiplicative) character of $\mathbb{F}_{p^d}$. For the quadratic form $f(x)$ defined by (\ref{eq:f(x)=Z(TAT')Z'}), we have
\[\sum_{x\in \mathbb{F}_q^s}\zeta_p^{{\rm Tr}_1^d(f(x))}=\left\{
  \begin{array}{ll}
    \eta_d(a_1\cdots a_r)q^{s-\frac{r}{2}}, & \hbox{if $q\equiv 1\ ({\rm mod}\ 4)$;} \\ \\
     (\sqrt{-1})^r\eta_d(a_1\cdots a_r)q^{s-\frac{r}{2}}, & \hbox{if $q\equiv 3\ ({\rm mod}\ 4)$.}
  \end{array}
\right.\]
\end{lemma}

By Definition~$2.1$, we have that $f_{\alpha,\beta}(x)={\rm Tr}_d^m(\alpha x^{p^k+1}+\beta x^2)$ is a quadratic form over $\mathbb{F}_{q}$. Let $r_{\alpha,\beta}$ denote the rank of $f_{\alpha,\beta}(x)$. Therefore,  $r_{\alpha,\beta}$ is determined by the codimension of \[ V=\{ x\in \mathbb{F}_{q^s} \mid f_{\alpha,\beta}(x+z)-f_{\alpha,\beta}(x)-f_{\alpha,\beta}(z)=0,\ \ {\rm for \ \ all }\ z\in \mathbb{F}_{q^s}\}, \]
which is determined by the number of solutions of
\begin{equation}\label{eq:phi(x)}
\phi_{\alpha,\beta}(x)=0,\ \ \ {\rm where} \ \ \phi_{\alpha,\beta}(x)=\alpha^{p^k}x^{p^{2k}}+2\beta^{p^k}x^{p^k}+\alpha x.
\end{equation}

Hence, we have the following result (see Lemma~$2$ \cite{LuoandFeng2}).
\begin{lemma}\label{lem:ralphabata}{\rm \cite{LuoandFeng2}}
For $(\alpha,\beta)\in \mathbb{F}_{p^m}^2\setminus \{(0,0)\},$ $r_{\alpha,\beta}$ is $s$, $s-1$ or $s-2$. Furthermore, let $n_i$ be the number of $f_{\alpha,\beta}(x)$ with  $r_{\alpha,\beta}=s-i$ for $(\alpha,\beta)\in \mathbb{F}_{p^m}^2\setminus \{(0,0)\}$ and $i=0,1,2$, then \[n_1=\frac{(p^m-1)(p^{m-1}-1)}{p^{2d}-1},\ n_2=p^{m-d}(p^m-1),\ n_0=p^{2m}-1-n_1-n_2.\]
\end{lemma}

We define
\begin{equation}\label{eq:Talphabata}
    T(\alpha,\beta)=\sum_{x\in \mathbb{F}_{p^m}}\zeta_p^{{\rm Tr}_1^m(\alpha x^{p^k+1}+\beta x^2)},\ \alpha,\beta \in \mathbb{F}_{p^m}.
\end{equation}
\begin{lemma}\label{lem:zhishuhefenbu}{\rm \cite{LuoandFeng2}}
With the notations given above. The value distribution of the exponential sum $T(\alpha,\beta)$ defined by (\ref{eq:Talphabata})
is shown in the following:

i) for the case $s$ is odd,
 \begin{center}
\begin{tabular}{|c|c|}
  \hline
  value & frequency \\ \hline
  $\pm\sqrt{q^*}q^{\frac{s-1}{2}}$,  & $\frac{1}{2}p^{2d}(p^m-p^{m-d}-p^{m-2d}+1)(p^m-1)/(p^{2d}-1)$ \\ \hline
  $p^{\frac{m+d}{2}}$ & $\frac{1}{2}p^{\frac{m-d}{2}}(p^{\frac{m-d}{2}}+1)(p^m-1)$ \\ \hline
  $-p^{\frac{m+d}{2}}$ & $\frac{1}{2}p^{\frac{m-d}{2}}(p^{\frac{m-d}{2}}-1)(p^m-1)$ \\ \hline
  $\pm\sqrt{q^*}q^{\frac{s+1}{2}}$ & $\frac{1}{2}(p^{m-d}-1)(p^m-1)/(p^{2d}-1)$ \\ \hline
  $p^m$ & $1$ \\ \hline
\end{tabular}
 \end{center}

ii) for the case $s$ is even,
\begin{center}
\begin{tabular}{|c|c|}
  \hline
  value & frequency \\ \hline
  $p^{\frac{m}{2}}$ & $\frac{1}{2}p^{2d}(p^m-p^{m-d}-p^{m-2d}+p^{\frac{m}{2}}-p^{\frac{m}{2}-d}+1)(p^m-1)/(p^{2d}-1)$ \\ \hline
    $-p^{\frac{m}{2}}$ & $\frac{1}{2}p^{2d}(p^m-p^{m-d}-p^{m-2d}-p^{\frac{m}{2}}+p^{\frac{m}{2}-d}+1)(p^m-1)/(p^{2d}-1)$ \\ \hline
  $\pm\sqrt{q^*}q^{\frac{s}{2}}$ & $\frac{1}{2}p^{m-d}(p^m-1)$ \\ \hline
  $p^{\frac{m}{2}+d}$ & $\frac{1}{2}(p^{\frac{m}{2}}-1)(p^{\frac{m}{2}-d}+1)(p^m-1)/(p^{2d}-1)$ \\ \hline
  $-p^{\frac{m}{2}+d}$ & $\frac{1}{2}(p^{\frac{m}{2}}+1)(p^{\frac{m}{2}-d}-1)(p^m-1)/(p^{2d}-1)$ \\ \hline
$p^m$ & $1$ \\ \hline
\end{tabular}
\end{center}
\end{lemma}
For convenience, we define $\varepsilon=\pm1$,  $i=0,1,2$,
$N_i=\{(\alpha,\beta)\in \mathbb{F}_{q}^2\mid r_{\alpha,\beta}=s-i\},$
\[N_{\varepsilon,i}=\left\{
                      \begin{array}{ll}
                        \{(\alpha,\beta)\in \mathbb{F}_{p^m}^2\setminus\{(0,0)\}\mid T(\alpha,\beta)=\varepsilon q_0^{\frac{s+i}{2}}\}, & \hbox{if $s+i$ is even;} \\
                        \{(\alpha,\beta)\in \mathbb{F}_{p^m}^2\setminus\{(0,0)\}\mid T(\alpha,\beta)=\varepsilon\sqrt{q_0^*} q_0^{\frac{s+i-1}{2}}\}, & \hbox{if $s+i$ is odd,}
                      \end{array}
                    \right.
\]
and $n_{\varepsilon,i}=\mid N_{\varepsilon,i}\mid$. Then, $N_i=N_{1,i}\cup N_{-1,i}$ and $n_i=n_{1,i}+n_{-1,i}$.
By (\ref{eq:phi(x)}), we define
\begin{equation}\label{eq:psix}
    \psi(\alpha,x)=-\frac{1}{2}x^{-1}(\alpha x^{p^k}+ \alpha^{p^{m-k}}x^{p^{m-k}}).
\end{equation}
Hence, $\beta=\psi(\alpha,x)$ if and only if $\phi_{\alpha,\beta}(x)=0$ with $x\in \mathbb{F}_q^*$. Therefore, it is easy to see that $r_{\alpha,\beta}=s-1$ or $s-2$ if and only if, $\alpha\in \mathbb{F}_{p^m}^*$ and there exists $x\in \mathbb{F}_{p^m}^*$ such that $\beta=\psi(\alpha,x)$. Furthermore, from Lemma~\ref{lem:ralphabata}, we obtain the number of solutions of $\beta=\psi(\alpha,x)$ with $x\in \mathbb{F}_{p^m}^*$ is
\begin{equation}\label{eq:alphabetafenbu}
    \#\{x\in \mathbb{F}_{p^m}^*\mid   \beta=\psi(\alpha,x)  \}=\left\{
                                                                 \begin{array}{ll}
                                                                   0, & \hbox{iff $(\alpha,\beta)\in N_0$;} \\
                                                                   p^d-1, & \hbox{iff $(\alpha,\beta)\in N_1$;} \\
                                                                   p^{2d}-1, & \hbox{iff $(\alpha,\beta)\in N_2$.}
                                                                 \end{array}
                                                               \right.
\end{equation}

At the end of this section, we investigate two classes of overdeterminate equations over finite fields.
\begin{lemma}\label{lem:E1}
Let $E_1$ denote the number of solutions $(x,y)\in \mathbb{F}_{p^m}^2$ of the following system equations
\begin{equation}\label{eq:x2+y2=0}
\left\{
  \begin{array}{ll}
    x^2+y^2=0,  \\
    x^{p^k+1}+y^{p^k+1}=0.
  \end{array}
\right.
\end{equation}
i) If $1\leq v_2(m)<v_2(k)$, then $E_1=2p^m-1$.\\
ii) If $ v_2(k)<v_2(m)$, then $E_1=2p^m-1$ for $p^k \equiv 1\ ({\rm mod}\ 4)$, $E_1=1$ for $p^k \equiv 3\ ({\rm mod}\ 4)$.
\end{lemma}
\pf By the first equation of (\ref{eq:x2+y2=0}), we have $-x^2=y^2$. Then
\begin{equation}\label{eq:sec1}
x^{p^k+1}+y^{p^k+1}=x^{p^k+1}+(-x^2)^{\frac{p^k+1}{2}}=\left\{
                                                           \begin{array}{ll}
                                                             0, & \hbox{if $p^k \equiv 1\ ({\rm mod}\ 4)$;} \\
                                                             2x^{p^k+1}, & \hbox{if $p^k \equiv 3\ ({\rm mod}\ 4)$.}
                                                           \end{array}
                                                         \right.
\end{equation}

If $1\leq v_2(m)<v_2(k)$, then $m$, $k$ are both even, which implies that $p^m \equiv 1\ ({\rm mod}\ 4)$ and $p^k \equiv 1\ ({\rm mod}\ 4)$. From (\ref{eq:sec1}) and the system equations (\ref{eq:x2+y2=0}), we have $E_1$ is equal to the number of solutions of equation $-x^2=y^2$, i.e., $x=\pm \pi^{\frac{p^m-1}{4}}y$. Then $E_1=2p^m-1$.

If $ v_2(k)<v_2(m)$,  we discuss it in two cases.

Case I, when $p^k \equiv 1\ ({\rm mod}\ 4)$: $E_1$ is equal to the number of solutions of equation $-x^2=y^2$ by (\ref{eq:x2+y2=0}) and (\ref{eq:sec1}). Note that $m$ is even, then $4\mid p^m-1$. So, $E_1=2p^m-1$.

Case II, when $p^k \equiv 3\ ({\rm mod}\ 4)$: From (\ref{eq:sec1}), we have $x^{p^k+1}+y^{p^k+1}=2x^{p^k+1}=0$, then $x=0$ and $y=0$. Hence, $E_1=1.$\EOP
\begin{lemma}\label{lem:E2}
Let $E_2$ denote the number of solutions $(x,y,z)\in \mathbb{F}_{p^m}^3$ of the following system equations
\begin{equation}\label{eq:x2+y2+z2=0}
\left\{
  \begin{array}{ll}
    x^2+y^2-\pi z^2=0,  \\
    x^{p^k+1}+y^{p^k+1}+\pi^{\frac{p^k+1}{2}} z^{p^k+1}=0.
  \end{array}
\right.
\end{equation}
If  $ v_2(k)<v_2(m)$, then $E_2=2p^m-1$ for $p^k \equiv 1\ ({\rm mod}\ 4)$, $E_2=1$ for $p^k \equiv 3\ ({\rm mod}\ 4)$.
\end{lemma}
\pf  We distinguish between the following two cases to calculate the number of solutions $(x,y,z)\in \mathbb{F}_{p^m}^3$ of (\ref{eq:x2+y2+z2=0}).

Case I, when $z=0$: If $z=0$, by Lemma~\ref{lem:E1} ii), we get that (\ref{eq:x2+y2+z2=0}) has $2p^m-1$ solutions for $p^k \equiv 1\ ({\rm mod}\ 4)$ or one solution for $p^k \equiv 3\ ({\rm mod}\ 4)$.

Case II, when $z\neq0$: It is easy to check that $x\neq0$ and $y\neq0$. In this case, one has that (\ref{eq:x2+y2+z2=0}) has  $(p^m-1)M$ solutions, where $M$ is the number of solutions of the following system equations
\begin{equation}\label{eq:x2+y2+z2=0dengjia}
\left\{
  \begin{array}{ll}
    x^2+y^2=\pi,  \\
    x^{p^k+1}+y^{p^k+1}=-\pi^{\frac{p^k+1}{2}}.
  \end{array}
\right.
\end{equation}
By (\ref{eq:x2+y2+z2=0dengjia}), we have $(x^2+y^2)^{p^k+1}=\pi^{p^k+1} $ and $(x^{p^k+1}+y^{p^k+1})^2=\pi^{p^k+1}$.  Combining these two equations (the first one minus the second one) leads to \[(xy^{p^k}-x^{p^k}y)^2=0 ,\] which shows that $x=\lambda y$, where $\lambda\in\mathbb{F}_{p^d}^*$ (since $\gcd(p^k-1,p^m-1)=p^d-1$). Substituting $x=\lambda y$ into  (\ref{eq:x2+y2+z2=0dengjia}), we get
\begin{equation}\label{eq:x2+y2+z2=0dengjia111}
\left\{
  \begin{array}{ll}
    (\lambda^2+1)y^2=\pi,  \\
    (\lambda^2+1)y^{p^k+1}=-\pi^{\frac{p^k+1}{2}}.
  \end{array}
\right.
\end{equation}
Eliminating $\lambda^2+1$ of (\ref{eq:x2+y2+z2=0dengjia111}), one has
\[ y^{p^k-1}=-\pi^{\frac{p^k-1}{2}}=\pi^{\frac{p^k-1}{2}+\frac{p^m-1}{2}},\]
which induces that $p^d-1\mid \frac{p^k-1}{2}+\frac{p^m-1}{2}$.
Since $v_2(k)<v_2(m)$, then $s=m/d$ is even. Hence, $p^d-1\mid \frac{p^m-1}{2}$. From  $v_2(k)<v_2(m)$, we have $k/d$ is odd and then $p^d-1\nmid \frac{p^k-1}{2}$. It is a contradiction with $p^d-1\mid \frac{p^k-1}{2}+\frac{p^m-1}{2}$.  So the number of solutions of (\ref{eq:x2+y2+z2=0dengjia111}) is $0$, i.e., $M=0$.

Therefore, $E_2=2p^m-1$ if $p^k \equiv 1\ ({\rm mod}\ 4)$, $E_2=1$ if $p^k \equiv 3\ ({\rm mod}\ 4)$.\EOP

\section{The weight distribution of  $\mathcal{C}$}
In this section, we first calculate the weight of the codeword $\textbf{c}(\alpha,\beta)\in \mathcal{C}$ defined by (\ref{eq:cycliccode}),
\begin{eqnarray}\label{eq:wtcab}
\nonumber  wt(\textbf{c}(\alpha,\beta))&= & \# \{0\leq i\leq p^m-2: c_i \neq0\} \\
\nonumber   &=& p^m-1-\frac{1}{p}\sum_{i=0}^{p^m-2}\sum_{u\in \mathbb{F}_p}\zeta_p^{u {\rm Tr}_1^m(\alpha\pi^{\frac{p^k+1}{2}i}+\beta(-\pi)^{i})} \\
\nonumber   &=& p^m-1-\frac{1}{p}\sum_{u\in \mathbb{F}_p}\sum_{i=0}^{\frac{p^m-3}{2}}\left(\zeta_p^{u {\rm Tr}_1^m(\alpha\pi^{\frac{p^k+1}{2}2i}+\beta\pi^{2i})}+\zeta_p^{u {\rm Tr}_1^m(\alpha\pi^{\frac{p^k+1}{2}(2i+1)}-\beta\pi^{(2i+1)})}\right) \\
\nonumber   &=& p^m-1-\frac{1}{p}\sum_{u\in \mathbb{F}_p}\sum_{x\in SQ}\left(\zeta_p^{u {\rm Tr}_1^m(\alpha x^{\frac{p^k+1}{2}}+\beta x)}+\zeta_p^{u {\rm Tr}_1^m(\alpha\pi^\frac{p^k+1}{2} x^{\frac{p^k+1}{2}}-\beta\pi x)}\right) \\
 \nonumber   &=& p^m-1-\frac{1}{2p}\sum_{u\in \mathbb{F}_p}\sum_{x\in \mathbb{F}_{p^m}^*}\left(\zeta_p^{u {\rm Tr}_1^m(\alpha x^{p^k+1}+\beta x^2)}+\zeta_p^{u {\rm Tr}_1^m(\alpha\pi^{\frac{p^k+1}{2}} x^{p^k+1}-\beta\pi x^2)}\right)\\
&=& p^{m}-p^{m-1}-\frac{1}{2p}\sum_{u\in \mathbb{F}_p^*}S(u\alpha,u\beta),
\end{eqnarray}
where
\begin{eqnarray}\label{eq:Salphabata}
\nonumber  S(\alpha,\beta) &=& \sum_{x\in \mathbb{F}_{p^m}}\left(\zeta_p^{ {\rm Tr}_1^m(\alpha x^{p^k+1}+\beta x^2)}+\zeta_p^{ {\rm Tr}_1^m(\alpha\pi^{\frac{p^k+1}{2}} x^{p^k+1}-\beta\pi x^2)}\right) \\
   &=& T(\alpha,\beta)+T(\pi^{\frac{p^k+1}{2}}\alpha ,-\pi\beta ),
\end{eqnarray}
where $T(\alpha,\beta)$ is defined by (\ref{eq:Talphabata}).

From (\ref{eq:wtcab}), the weight distribution of  $\mathcal{C}$ is completely
determined by the value distribution of $S(\alpha,\beta)$.
To obtain the value distribution of $S(\alpha,\beta)$, we need a series of lemmas. Before introducing them, we define $f(x)= {\rm Tr}_d^m(\alpha x^{p^k+1}+\beta x^2)$ and $g(x)={ {\rm Tr}_d^m(\pi^{\frac{p^k+1}{2}}\alpha x^{p^k+1}-\pi\beta  x^2)}$. Let $r_f$ and $r_g$ be the rank of $f(x)$ and $g(x)$, respectively. Then, we have the following result.
\begin{lemma}\label{lem:rfrg}
For $(\alpha,\beta)\in \mathbb{F}_{p^m}^2 \setminus \{(0,0)\}$, when $ 1\leq v_2(m)<v_2(k)$ or $ v_2(k)<v_2(m)$,  we have that at least one of $r_f$ and $f_g$ is $s$.
\end{lemma}
\pf Assume that $r_f$ and $f_g$ are both less than $s$. By (\ref{eq:phi(x)}), we have that there exists $x_1, x_2\in \mathbb{F}_{p^m}^*$ such that
\begin{equation}\label{eq:x1}
\phi_{\alpha,\beta}(x_1)=\alpha^{p^k}x_1^{p^{2k}}+2\beta^{p^k}x_1^{p^k}+\alpha x_1=0,
\end{equation}
and
\begin{equation}\label{eq:x2}
\phi_{\pi^{(p^k+1)/2}\alpha,-\pi\beta}(x_2)=(\pi^{\frac{p^k+1}{2}}\alpha)^{p^k}x_2^{p^{2k}}-2(\pi\beta)^{p^k}x_2^{p^k}+\pi^{\frac{p^k+1}{2}}\alpha x_2=0.
\end{equation}
If $\alpha=0$, we have $x_1=x_2=0$, which is a contradiction with $x_1, x_2\in \mathbb{F}_{p^m}^*$. In the following, we assume that $\alpha\neq 0$.
Simplifying (\ref{eq:x1})$\times (\pi x_2)^{p^k}+$(\ref{eq:x2})$\times  x_1^{p^k}$, one has
\begin{equation}\label{eq:maodunshizi}
-\pi^{\frac{p^k-1}{2}}\left(\alpha x_1 x_2(x_1^{p^k-1}+\pi^{\frac{p^k-1}{2}} x_2^{p^k-1})\right)^{p^k}=\alpha x_1 x_2(x_1^{p^k-1}+\pi^{\frac{p^k-1}{2}} x_2^{p^k-1}),
\end{equation}
which implies that $$x_1^{p^k-1}+\pi^{\frac{p^k-1}{2}} x_2^{p^k-1}=0$$ or
\[\left(\alpha x_1 x_2(x_1^{p^k-1}+\pi^{\frac{p^k-1}{2}} x_2^{p^k-1})\right)^{p^k-1}=-\pi^{-\frac{p^k-1}{2}}. \]
If one of the above two equations holds, we have that $p^d-1\mid \frac{p^m-1}{2}-\frac{p^k-1}{2}$.

Case I, when $ 1\leq v_2(m)<v_2(k)$: In this case, $s=m/d$ is odd and  we get that $p^d-1\nmid \frac{p^m-1}{2}$. From $1\leq v_2(m)<v_2(k)$, we have that $k/d$ is even. This implies that  $p^d-1\mid \frac{p^k-1}{2}$ and then it is a contradiction with $p^d-1\mid \frac{p^m-1}{2}-\frac{p^k-1}{2}$.

Case II, $ v_2(k)<v_2(m)$: Similarly, we can prove that $p^d-1\nmid \frac{p^m-1}{2}-\frac{p^k-1}{2}$, which is also a contradiction with $p^d-1\mid \frac{p^m-1}{2}-\frac{p^k-1}{2}$.

Therefore, we have that at least one of $r_f$ and $f_g$ is $s$. \EOP

\subsection{The weight distribution of $\mathcal{C}$ for $1\leq v_2(m)<v_2(k)$}
In this subsection, we always assume that $1\leq v_2(m)<v_2(k)$.
To determine the value distribution of $S(\alpha,\beta)$, we  need the following lemma.
\begin{lemma}\label{lem:2gexingzhi}
With the notations given above, we have the following result:
\begin{enumerate}
  \item $\sum\limits_{\alpha,\beta\in \mathbb{F}_{p^m}}S(\alpha,\beta)^2=                                  4 p^{3m}.$
  \item $(p^d-1)\sum\limits_{(\alpha,\beta)\in N_1}S(\alpha,\beta)^2+(p^{2d}-1)\sum\limits_{(\alpha,\beta)\in N_2}S(\alpha,\beta)^2=p^{m}(p^m-1)(2p^{m+d}-2p^m+2p^d-p^{2d}-1).$
\end{enumerate}
\end{lemma}
\pf 1. By (\ref{eq:Salphabata}), we get that
\begin{equation}\label{eq:S^2+}
\sum_{\alpha,\beta\in \mathbb{F}_{p^m}}S(\alpha,\beta)^2
=\sum_{\alpha,\beta\in\mathbb{F}_{p^m}}\left(T(\alpha,\beta)^2+2T(\alpha,\beta)T(\pi^{\frac{p^k+1}{2}}\alpha,-\pi \beta)+ T(\pi^{\frac{p^k+1}{2}}\alpha,-\pi \beta)^2\right).
\end{equation}
On one hand, by (\ref{eq:Talphabata}), we have
\begin{eqnarray}\label{eq:Mp2m+}
\nonumber \sum_{\alpha,\beta\in\mathbb{F}_{p^m}}T(\alpha,\beta)^2 &=& \sum_{x,y\in\mathbb{F}_{p^m}}\sum_{\alpha\in\mathbb{F}_{p^m}}\zeta_p^{ {\rm Tr}_1^m(\alpha (x^{p^k+1}+y^{p^k+1}))}\sum_{\beta\in\mathbb{F}_{p^m}}\zeta_p^{ {\rm Tr}_1^m(\alpha (x^{2}+y^{2}))} \\
 &=& Mp^{2m},
\end{eqnarray}
where $M=\#\{(x,y)\in \mathbb{F}_{p^m}^2\mid x^2+y^2=0, x^{p^k+1}+y^{p^k+1}=0\}$. By Lemma~\ref{lem:E1} i) and  (\ref{eq:Mp2m+}),
\[\sum_{\alpha,\beta\in\mathbb{F}_{p^m}}T(\alpha,\beta)^2=p^{2m}(2p^m-1).\]
Similarly, \[\sum_{\alpha,\beta\in\mathbb{F}_{p^m}}T(\pi^{\frac{p^k+1}{2}}\alpha,-\pi \beta)^2= p^{2m}(2p^m-1). \]
On the other hand, by (\ref{eq:Talphabata}),
\begin{eqnarray*}
   && \sum_{\alpha,\beta\in\mathbb{F}_{p^m}}T(\alpha,\beta)T(\pi^{\frac{p^k+1}{2}}\alpha,-\pi \beta) \\
   &=& \sum_{x,y\in\mathbb{F}_{p^m}}\sum_{\alpha\in\mathbb{F}_{p^m}}\zeta_p^{ {\rm Tr}_1^m(\alpha( x^{p^k+1}+\pi^{\frac{p^k+1}{2}}y^{p^k+1}))}\sum_{\beta\in\mathbb{F}_{p^m}}\zeta_p^{ {\rm Tr}_1^m(\beta (x^{2}-\pi y^{2}))}  \\
   &=&  \#\{(x,y)\in \mathbb{F}_{p^m}^2\mid x^2-\pi y^2=0, x^{p^k+1}+\pi^{\frac{p^k+1}{2}}y^{p^k+1}=0\}\cdot p^{2m}\\
   &=&  \#\{(x,y)\in \mathbb{F}_{p^m}^2\mid x^2-\pi y^2=0, 2\pi^{\frac{p^k+1}{2}}y^{p^k+1}=0\}\cdot p^{2m}\\
&=& p^{2m}.
\end{eqnarray*}
Hence, from (\ref{eq:S^2+}),
\[\sum_{\alpha,\beta\in \mathbb{F}_{p^m}}S(\alpha,\beta)^2 =4 p^{3m}. \]

2. By (\ref{eq:alphabetafenbu}) and (\ref{eq:Salphabata}), we have that
\begin{eqnarray}\label{eq:11-}
\nonumber   && (p^d-1)\sum\limits_{(\alpha,\beta)\in N_1}S(\alpha,\beta)^2+(p^{2d}-1)\sum\limits_{(\alpha,\beta)\in N_2}S(\alpha,\beta)^2 \\
&=&  \sum_{x,\alpha \in \mathbb{F}_{p^m}^*}S(\alpha,\psi(\alpha,x))^2
= \sum_{x,\alpha \in \mathbb{F}_{p^m}^*}\left(T(\alpha,\psi(\alpha,x))+T(\pi^{\frac{p^k+1}{2}}\alpha,-\pi \psi(\alpha,x))\right)^2.
\end{eqnarray}
By (\ref{eq:Talphabata}) and (\ref{eq:psix}),
\begin{eqnarray}\label{eq:12-}
\nonumber   & & \sum_{x,\alpha \in \mathbb{F}_{p^m}^*}T(\alpha,\psi(\alpha,x))^2\\
\nonumber  &=& \sum_{x,\alpha \in \mathbb{F}_{p^m}^*}\sum_{y,z\in \mathbb{F}_{p^m}}\zeta_p^{ {\rm Tr}_1^m(\alpha (y^{p^k+1}+y^{p^k+1}) -\frac{1}{2}(\alpha x^{p^k-1}+ \alpha^{p^{m-k}}x^{p^{m-k}-1})(y^2+z^2))}\\
\nonumber   &=& \sum_{x,\alpha \in \mathbb{F}_{p^m}^*}\sum_{y,z\in \mathbb{F}_{p^m}}\zeta_p^{ {\rm Tr}_1^m(\alpha (y^{p^k+1}+z^{p^k+1}) -\frac{1}{2}\alpha x^{p^k-1}(y^2+z^2)-\frac{1}{2}x^{1-p^k} \alpha (y^{2p^k}+z^{2p^k}))}\\
\nonumber&=& \sum_{x\in\mathbb{F}_{p^m}^*,y,z\in\mathbb{F}_{p^m}}\sum_{\alpha \in \mathbb{F}_{p^m}^* }\zeta_p^{ {\rm Tr}_1^m(-\frac{1}{2}\alpha x^{1-p^k}(( x^{p^k-1}y-y^{p^k})^2+( x^{p^k-1}z-z^{p^k})^2))}\\
\nonumber&=& (p^m-1)M_1-(p^{2m}(p^m-1)-M_1)\\
&=&p^mM_1-p^{3m}+p^{2m},
\end{eqnarray}
where
\[M_1=\#\{x\in \mathbb{F}_{p^m}^*,y,z\in \mathbb{F}_{p^m}\mid ( x^{p^k-1}y-y^{p^k})^2+( x^{p^k-1}z-z^{p^k})^2=0\}.
\]
For a fixed $x\in \mathbb{F}_{p^m}^*$, we  investigate the following equation
\begin{equation}\label{eq:13+}
    ( x^{p^k-1}y-y^{p^k})^2=-( x^{p^k-1}z-z^{p^k})^2.
\end{equation}
Since  $m$ is even, then $4\mid p^m-1$.    This shows that $-1=\pi ^{\frac{p^m-1}{2}}$ is a square element in $\mathbb{F}_{p^m}$.

In the following, we discuss it case by case.

Case I, when $y=0$: In this case, we have  $x^{p^k-1}z-z^{p^k}=0$, i.e., $z=\lambda x$, where $\lambda\in \mathbb{F}_{p^d}$. Hence, (\ref{eq:13+}) has $p^d$ solutions.

Case II, when $z=0$ and $y\neq 0$: We get that $y=tx$, where $t\in \mathbb{F}_{p^d}^*$, which implies that (\ref{eq:13+}) has $p^d-1$ solutions.

Case III, when $yz\neq 0$: Since $k$ is even, then $(\pi ^{\frac{p^m-1}{4}})^{p^k}=\pi ^{\frac{p^m-1}{4}}$. Together with $-1=\pi ^{\frac{p^m-1}{2}}$ and by (\ref{eq:13+}), we have
\[ y^{p^k}-yx^{p^k-1}=\pm \pi ^{\frac{p^m-1}{4}}(z^{p^k}-zx^{p^k-1} ),\]
which is equivalent to
\begin{equation}\label{eq:liangge1}
(y-\pi ^{\frac{p^m-1}{4}}z)^{p^k}=(y-\pi ^{\frac{p^m-1}{4}}z)x^{p^k-1}
\end{equation}
 or
\begin{equation}\label{eq:liangge2}
(y+\pi ^{\frac{p^m-1}{4}}z)^{p^k}=(y+\pi ^{\frac{p^m-1}{4}}z)x^{p^k-1}.
\end{equation}
Since $\gcd(p^k-1,p^m-1)=p^d-1$ and from (\ref{eq:liangge1}) and (\ref{eq:liangge2}), we get $y-\pi ^{\frac{p^m-1}{4}}z=rx$ and $y+\pi ^{\frac{p^m-1}{4}}z=r'x$, respectively, where $r,r'\in \mathbb{F}_{p^d}$. Hence, (\ref{eq:liangge1}) and (\ref{eq:liangge2}) have $p^{m+d}$  solutions, respectively. Note that the common solutions of (\ref{eq:liangge1}) and (\ref{eq:liangge2}) are $p^{2d}$. Therefore, the total number of pairs $(y,z)\in \mathbb{F}_{p^d}^2$ satisfying (\ref{eq:liangge1}) and (\ref{eq:liangge2}) is $2p^{m+d}-p^{2d}$. In which the number satisfying $yz=0$ is $2p^d-1$. Then, when $yz\neq 0$, the total number of pairs $(y,z)\in \mathbb{F}_{p^d}^2$ satisfying (\ref{eq:liangge1}) and (\ref{eq:liangge2}) is $2p^{m+d}-p^{2d}-2p^d+1$.

As a result, we have
\[M_1=(p^d+p^d-1+2p^{m+d}-p^{2d}-2p^d+1)(p^m-1)=p^d(2p^m-p^d)(p^m-1).\]

Similarly, by (\ref{eq:Talphabata}),
\begin{eqnarray}\label{eq:13-}
\nonumber   & & \sum_{x,\alpha \in \mathbb{F}_{p^m}^*}T(\alpha,\psi(\alpha,x))T(\pi^{\frac{p^k+1}{2}}\alpha,-\pi \psi(\alpha,x))\\
\nonumber   &=&  \sum_{x,\alpha \in \mathbb{F}_{p^m}^*}\sum_{y,z\in \mathbb{F}_{p^m}}\zeta_p^{ {\rm Tr}_1^m(\alpha (y^{p^k+1}+\pi^{\frac{p^k+1}{2}}z^{p^k+1}) -\frac{1}{2}(\alpha x^{p^k-1}+ \alpha^{p^{m-k}}x^{p^{m-k}-1})(y^2-\pi z^2))}\\
\nonumber   &=&  \sum_{x\in\mathbb{F}_{p^m}^*,y,z\in\mathbb{F}_{p^m}}\sum_{\alpha \in \mathbb{F}_{p^m}^* }\zeta_p^{ {\rm Tr}_1^m(\frac{1}{2}\alpha x^{1-p^k}(-( x^{p^k-1}y-y^{p^k})^2+\pi( x^{p^k-1}z+\pi^{\frac{p^k-1}{2}}z^{p^k})^2))}\\
\nonumber&=& (p^m-1)M_2-(p^{2m}(p^m-1)-M_2)\\
&=&p^mM_2-p^{3m}+p^{2m},
\end{eqnarray}
where
\[M_2=\#\{x\in \mathbb{F}_{p^m}^*,y,z\in \mathbb{F}_{p^m}\mid ( x^{p^k-1}y-y^{p^k})^2=\pi( x^{p^k-1}z+\pi^{\frac{p^k-1}{2}}z^{p^k})^2\}.
\]
For a given $x\in \mathbb{F}_{p^m}^*$, we  study the following equation

\begin{equation}\label{eq:44}
( x^{p^k-1}y-y^{p^k})^2=\pi( x^{p^k-1}z+\pi^{\frac{p^k-1}{2}}z^{p^k})^2.
\end{equation}

Case I, when $y=0$: In this case, we get that $z=0$ or $x^{p^k-1}=-\pi^{\frac{p^k-1}{2}}z^{p^k-1} $. Note that $s$ is odd, then $p^d-1 \nmid \frac{p^m-1}{2}$. Since $\frac{k}{d}$ is even, then $p^d-1 \mid \frac{p^k-1}{2}$. Hence, there is no solution of $x^{p^k-1}=-\pi^{\frac{p^k-1}{2}}z^{p^k-1}$ for $z\in \mathbb{F}_{p^d}^*$. In this case, (\ref{eq:44}) has only one solution.

Case II, when $z=0$ and $y\neq 0$: We have that $x^{p^k-1}y-y^{p^k}=0$, which is equivalent to $x^{p^k-1}=y^{p^k-1}$, i.e., $y=tx$, where $t\in \mathbb{F}_{p^d}^*$. Hence, (\ref{eq:44}) has $p^d-1$ solutions.

Case III, when $yz\neq 0$:   From case I, we get that $x^{p^k-1}+\pi^{\frac{p^k-1}{2}}z^{p^k-1}\neq 0$ for $z\neq 0$, which implies that  $x^{p^k-1}y-y^{p^k}\neq 0$ by (\ref{eq:44}). Then  there is no solution of (\ref{eq:44}).

Therefore, $M_2=p^d(p^m-1)$.

Similarly, by (\ref{eq:Talphabata}),
\begin{eqnarray}\label{eq:14-}
\nonumber   & & \sum_{x,\alpha \in \mathbb{F}_{p^m}^*}T(\pi^{\frac{p^k+1}{2}}\alpha,-\pi \psi(\alpha,x))^2\\
\nonumber   &=&  \sum_{x,\alpha \in \mathbb{F}_{p^m}^*}\sum_{y,z\in \mathbb{F}_{p^m}}\zeta_p^{ {\rm Tr}_1^m(\alpha\pi^{\frac{p^k+1}{2} } (y^{p^k+1}+z^{p^k+1}) +\frac{1}{2}\pi(\alpha x^{p^k-1}+ \alpha^{p^{m-k}}x^{p^{m-k}-1})(y^2+z^2))}\\
\nonumber   &=&  \sum_{x\in\mathbb{F}_{p^m}^*,y,z\in\mathbb{F}_{p^m}}\sum_{\alpha \in \mathbb{F}_{p^m}^* }\zeta_p^{ {\rm Tr}_1^m(\frac{1}{2}\pi\alpha x^{1-p^k}(( x^{p^k-1}y+\pi^{\frac{p^k-1}{2} }y^{p^k})^2+( x^{p^k-1}z+\pi^{\frac{p^k-1}{2}}z^{p^k})^2))}\\
\nonumber&=& (p^m-1)M_3-(p^{2m}(p^m-1)-M_3)\\
&=&p^mM_3-p^{3m}+p^{2m},
\end{eqnarray}
where
\[M_3=\#\{x\in \mathbb{F}_{p^m}^*,y,z\in \mathbb{F}_{p^m}\mid ( x^{p^k-1}y+\pi^{\frac{p^k-1}{2} }y^{p^k})^2+( x^{p^k-1}z+\pi^{\frac{p^k-1}{2}}z^{p^k})^2=0\}.
\]
For a given $x\in \mathbb{F}_{p^m}^*$, we  investigate the following equation
\[
( x^{p^k-1}y+\pi^{\frac{p^k-1}{2} }y^{p^k})^2=-( x^{p^k-1}z+\pi^{\frac{p^k-1}{2}}z^{p^k})^2,
\]
which is equivalent to
\begin{equation}\label{eq:liangge3}
-\pi^{\frac{p^k-1}{2}}(y-\pi ^{\frac{p^m-1}{4}}z)^{p^k}=(y-\pi ^{\frac{p^m-1}{4}}z)x^{p^k-1}
\end{equation}
 or
\begin{equation}\label{eq:liangge4}
-\pi^{\frac{p^k-1}{2}}(y+\pi ^{\frac{p^m-1}{4}}z)^{p^k}=(y+\pi ^{\frac{p^m-1}{4}}z)x^{p^k-1}.
\end{equation}

Case I, when $y-\pi ^{\frac{p^m-1}{4}}z=0$ and $y+\pi ^{\frac{p^m-1}{4}}z=0$: In this case, $y=z=0$. (\ref{eq:liangge3}) has one solutions and (\ref{eq:liangge4}) has one solution, respectively.

Case II, when $y-\pi ^{\frac{p^m-1}{4}}z=0$ and $y+\pi ^{\frac{p^m-1}{4}}z\neq 0$:
(\ref{eq:liangge4}) is equivalent to $-\pi^{\frac{p^k-1}{2}}(y+\pi ^{\frac{p^m-1}{4}}z)^{p^k-1}=x^{p^k-1}.$ Note that $m/d$ is odd and $k/d$ is even, then $p^d-1\nmid \frac{p^m-1}{2} $ and $p^d-1\mid \frac{p^k-1}{2} $. This implies that there is no solution of $-\pi^{\frac{p^k-1}{2}}(y+\pi ^{\frac{p^m-1}{4}}z)^{p^k-1}=x^{p^k-1}.$ Hence, (\ref{eq:liangge3}) has $p^m-1$ solutions, (\ref{eq:liangge4}) has no solution.

Case III, when $y-\pi ^{\frac{p^m-1}{4}}z\neq0$ and $y+\pi ^{\frac{p^m-1}{4}}z= 0$: We can discuss it by a similar way in the case II. Hence, (\ref{eq:liangge3}) has no solution and (\ref{eq:liangge4}) has $p^m-1$ solutions, respectively.

Case IV, when $(y-\pi ^{\frac{p^m-1}{4}}z)(y+\pi ^{\frac{p^m-1}{4}}z)\neq0$: Similarly, we have that (\ref{eq:liangge3}) and (\ref{eq:liangge4}) have no solution, respectively.

So, we get that $M_3=(2p^m-1)(p^m-1)$.

From (\ref{eq:11-}), (\ref{eq:12-}),  (\ref{eq:13-}) and (\ref{eq:14-}), we finish the proof.\EOP\\


In the following, we define
$$N_{\varepsilon,\mu,i,j}=\{(\alpha,\beta)\mid (\alpha,\beta)\in N_{\varepsilon,i}, (\pi^{\frac{p^k+1}{2}}\alpha,-\pi\beta)\in N_{\mu,j} \} \ \ {\rm  and} \ \ n_{\varepsilon,\mu,i,j}=\mid N_{\varepsilon,\mu,i,j} \mid $$ for $\varepsilon,\mu=\pm 1$, $0\leq i,j \leq 2$.

By Lemma~\ref{lem:rfrg}, we have that if $ij\neq0$, then $n_{\varepsilon,\mu,i,j}=0$. Moreover, it is easy to check that $n_{\varepsilon,\mu,i,j}=n_{\mu,\varepsilon,j,i}$ by the symmetry of $T(\alpha,\beta)$ and $T(\pi^{\frac{p^k+1}{2}}\alpha,-\pi\beta)$. Hence, we need only
to calculate $n_{\varepsilon,\mu,i,0}.$ For convenience, we let $n_{\varepsilon,\mu,i,0}=n_{\varepsilon,\mu,i}$.
With above preparation we can determine the value distribution of the exponential sum $S(\alpha,\beta)$.

\begin{theorem}\label{th:Svaluedistribution}
With the notations given above. Then the value distribution of $ S(\alpha,\beta)$ defined by (\ref{eq:Salphabata}) is as follows:
\begin{center}
\begin{tabular}{|l|l|}
  \hline
  value & frequency \\ \hline
  $2p^m$ & $1$ \\ \hline
  $0$ & $\frac{1}{2}(p^{m+d}-3p^m+p^d+1)(p^m-1)/(p^d-1)$ \\ \hline
  $\pm (p^d-1)p^{\frac{m}{2}}$ & $(p^{m-d}-1)(p^m-1)/(p^{2d}-1)$ \\ \hline
  $ (\pm1 -p^{\frac{d}{2}})p^{\frac{m}{2}}$ & $\frac{1}{2}p^{\frac{m-d}{2}}(p^{\frac{m-d}{2}}-1)(p^m-1)$ \\ \hline
  $ (\pm1 +p^{\frac{d}{2}})p^{\frac{m}{2}}$ & $\frac{1}{2}p^{\frac{m-d}{2}}(p^{\frac{m-d}{2}}+1)(p^m-1)$ \\ \hline
  $\pm2p^{\frac{m}{2}}$  & $\frac{1}{4}(p^d-1)(p^{2m}-1)/(p^d+1)$ \\
  \hline
\end{tabular}
\end{center}
\end{theorem}

\pf From Lemma~\ref{lem:rfrg} and Lemma~\ref{lem:zhishuhefenbu}, we get that
\begin{equation}\label{eq:th1}
    \left\{
      \begin{array}{ll}
        n_{1,0}=n_{1,1,0}+n_{-1,1,0}+n_{1,1,1}+n_{-1,1,1}+n_{1,1,2}+n_{-1,1,2},  \\
        n_{-1,0}=n_{1,-1,0}+n_{-1,-1,0}+n_{1,-1,1}+n_{-1,-1,1}+n_{1,-1,2}+n_{-1,-1,2}.
      \end{array}
    \right.
\end{equation}
By (\ref{eq:Salphabata}) and Lemma~\ref{lem:rfrg}, we have that
\begin{equation}\label{eq:th2}
    \left\{
      \begin{array}{ll}
        n_{1,1}=n_{1,1,1}+n_{1,-1,1},  \\
        n_{-1,1}=n_{-1,1,1}+n_{-1,-1,1},\\
        n_{1,2}=n_{1,1,2}+n_{1,-1,2},  \\
        n_{-1,2}=n_{-1,1,2}+n_{-1,-1,2}.
      \end{array}
    \right.
\end{equation}
Note that if the rank of ${\rm Tr}_d^m(\alpha x^{p^k+1}+\beta x^2)$  is odd and we compute the value of $T(\alpha,\beta)$ and $T(a\alpha,a\beta)$, where $a$ is a nonsquare element in $\mathbb{F}_q$. By Lemma~\ref{lem:erciquzhi}, we have
\[T(a\alpha,a\beta)=-T(\alpha,\beta), \]
and since $s$ is odd, we have
\begin{equation}\label{eq:th3}
    \left\{
      \begin{array}{ll}
        n_{1,1,0}=n_{-1,-1,0},  \\
        n_{1,-1,0}=n_{-1,1,0},\\
 n_{1,1,1}=n_{1,-1,1},  \\
        n_{-1,1,1}=n_{-1,-1,1},\\
 n_{1,1,2}=n_{-1,-1,2},  \\
        n_{1,-1,2}=n_{-1,1,2}.
      \end{array}
    \right.
\end{equation}
By Lemma~\ref{lem:2gexingzhi} 1), we get that
\begin{equation}\label{eq:lianggefangcheng1}
  \begin{array}{ll}
    4p^{3m}=&4p^{2m}+4p^m(n_{1,1,0}+n_{-1,-1,0})+2(p^{\frac{d}{2}}+1)^2p^{m}(n_{1,1,1}+n_{-1,-1,1})  \\
    &+2(p^{\frac{d}{2}}-1)^2p^{m}(n_{1,-1,1}+n_{-1,1,1})+2(p^d+1)^2p^m(n_{1,1,2}+n_{-1,-1,2})\\
&+2(p^d-1)^2p^m(n_{1,-1,2}+n_{-1,1,2}).
  \end{array}
\end{equation}
By Lemma~\ref{lem:2gexingzhi} 2), we have that
\begin{equation}\label{eq:lianggefangcheng2}
  \begin{array}{ll}
&p^{m}(p^m-1)(2p^{m+d}-2p^m+2p^d-p^{2d}-1)\\
=&(p^d-1)p^{m}((p^{\frac{d}{2}}+1)^2(n_{1,1,1}+n_{-1,-1,1})+(p^{\frac{d}{2}}-1)^2(n_{1,-1,1}+n_{-1,1,1})\\
&+(p^d+1)^3(n_{1,1,2}+n_{-1,-1,2})+(p^d+1)(p^d-1)^2(n_{1,-1,2}+n_{-1,1,2})).
\end{array}
\end{equation}
Solving the system of equations consisting of (\ref{eq:th1})-(\ref{eq:lianggefangcheng2}), we have
\[n_{1,1,0}=n_{-1,-1,0}=\frac{1}{4}(p^d-1)(p^{2m}-1)/(p^d+1),\]
\[n_{1,-1,0}=n_{-1,1,0}=\frac{1}{4}(p^{m+d}-3p^m+p^d+1)(p^m-1)/(p^d-1),\]
\[n_{-1,1,1}=n_{-1,-1,1}=\frac{1}{4}p^{\frac{m-d}{2}}(p^{\frac{m-d}{2}}-1)(p^m-1),\]
\[n_{1,1,1}=n_{1,-1,1}=\frac{1}{4}p^{\frac{m-d}{2}}(p^{\frac{m-d}{2}}+1)(p^m-1),\]
\[n_{1,-1,2}=n_{-1,1,2}=\frac{1}{2}(p^{m-d}-1)(p^m-1)/(p^{2d}-1),\]
\[n_{1,1,2}=n_{-1,-1,2}=0. \]
We complete the proof.\EOP

Therefore, we can give the weight distribution of cyclic code $\mathcal{C}$.
\begin{theorem}\label{th:Cdistribution}
With the notations given above. Then $\mathcal{C}$ defined by (\ref{eq:cycliccode}) is a cyclic code over $\mathbb{F}_p$ with length $p^m-1$ and dimension $2m$. Moreover, the weight distribution of cyclic code $ \mathcal{C}$  is given in Table~$1$.
\begin{table}[!h]
\tabcolsep 0pt
\caption{weight distribution of $ \mathcal{C}$ for $1\leq v_2(m)<v_2(k)$}
\vspace*{-12pt}
\begin{center}
\def\temptablewidth{0.9\textwidth}
{\rule{\temptablewidth}{1pt}}
\begin{tabular*}{\temptablewidth}{@{\extracolsep{\fill}}cc}
Weight & Frequency \\   \hline
  $0$ & $1$ \\
  $p^{m-1}(p-1)+ \frac{1}{2}(p-1)(p^d-1)p^{\frac{m}{2}-1}$ & $(p^{m-d}-1)(p^m-1)/(p^{2d}-1)$ \\
$p^{m-1}(p-1)- \frac{1}{2}(p-1)(p^d-1)p^{\frac{m}{2}-1}$ & $(p^{m-d}-1)(p^m-1)/(p^{2d}-1)$ \\
  $p^{m-1}(p-1)+ \frac{1}{2}(p-1)(p^{\frac{d}{2}}-1)p^{\frac{m}{2}-1}$ & $\frac{1}{2}p^{\frac{m-d}{2}}(p^{\frac{m-d}{2}}-1)(p^m-1)$ \\
$p^{m-1}(p-1) +\frac{1}{2}(p-1)(p^{\frac{d}{2}}+1)p^{\frac{m}{2}-1}$ & $\frac{1}{2}p^{\frac{m-d}{2}}(p^{\frac{m-d}{2}}-1)(p^m-1)$ \\
  $p^{m-1}(p-1)- \frac{1}{2}(p-1)(p^{\frac{d}{2}}+1)p^{\frac{m}{2}-1}$ & $\frac{1}{2}p^{\frac{m-d}{2}}(p^{\frac{m-d}{2}}+1)(p^m-1)$ \\
$p^{m-1}(p-1) -\frac{1}{2}(p-1)(p^{\frac{d}{2}}-1)p^{\frac{m}{2}-1}$ & $\frac{1}{2}p^{\frac{m-d}{2}}(p^{\frac{m-d}{2}}+1)(p^m-1)$ \\
  $p^{m-1}(p-1)+p^{\frac{m}{2}-1}(p-1)$  & $\frac{1}{4}(p^d-1)(p^{2m}-1)/(p^d+1)$ \\
$p^{m-1}(p-1)-p^{\frac{m}{2}-1}(p-1)$  & $\frac{1}{4}(p^d-1)(p^{2m}-1)/(p^d+1)$ \\
$p^{m-1}(p-1)$ & $\frac{1}{2}(p^{m+d}-3p^m+p^d+1)(p^m-1)/(p^d-1)$
\end{tabular*}
{\rule{\temptablewidth}{1pt}}
\end{center}
\end{table}
\end{theorem}
\pf By (\ref{eq:wtcab}), we have
\[wt(\textbf{c}(\alpha,\beta))=p^m-p^{m-1}-\frac{1}{2p}\sum_{u\in \mathbb{F}_p^*}S(u\alpha,u\beta).\]
Note that $m$ is even, then $u_p=\pi^{\frac{p^m-1}{p-1}}=\pi^{p^{m-1}+\cdots+1}$ is a square element in $\mathbb{F}_{p^m}$. For a given $u\in\mathbb{F}_{p}^*$, we have that $u\in SQ$. Since $k$ is even, then $u^{\frac{p^k+1}{2}}=u$. So we have
\begin{eqnarray*}
  S(u\alpha,u\beta)
   &=&  \sum_{x\in \mathbb{F}_{p^m}}\left(\zeta_p^{ {\rm Tr}_1^m(u\alpha x^{p^k+1}+u\beta x^2)}+\zeta_p^{ {\rm Tr}_1^m(u\alpha\pi^{\frac{p^k+1}{2}} x^{p^k+1}-u\beta\pi x^2)}\right)\\
&=& \sum_{x\in \mathbb{F}_{p^m}}\left(\zeta_p^{ {\rm Tr}_1^m(\alpha (u^{\frac{1}{2}}x)^{p^k+1}+\beta (u^{\frac{1}{2}}x)^2)}+\zeta_p^{ {\rm Tr}_1^m(\alpha\pi^{\frac{p^k+1}{2}} (u^{\frac{1}{2}}x)^{p^k+1}-\beta\pi (u^{\frac{1}{2}}x)^2)}\right)\\
&=&  \sum_{x\in \mathbb{F}_{p^m}}\left(\zeta_p^{ {\rm Tr}_1^m(\alpha x^{p^k+1}+\beta x^2)}+\zeta_p^{ {\rm Tr}_1^m(\alpha\pi^{\frac{p^k+1}{2}} x^{p^k+1}-\beta\pi x^2)}\right)\\
&=& S(\alpha,\beta).
\end{eqnarray*}
Therefore, we obtain that
\[wt(\textbf{c}(\alpha,\beta))=p^m-p^{m-1}-\frac{p-1}{2p}S(\alpha,\beta).\]
By Theorem~\ref{th:Svaluedistribution}, we get the result.\EOP

In the following, we give an example to verify our  result in Theorem~\ref{th:Cdistribution}.
\begin{example}
Let $p=3$, $m=6$, $k=4$, the code $\mathcal{C}$ is a $[728,12,414]$ cyclic
code over $\mathbb{F}_3$ with weight enumerator
\[1+728X^{414}+32760X^{450}+139048X^{468}+199472X^{486}+132496X^{504}+26208X^{522}+728X^{558},\]
which confirms the weight distribution in Table $1$.
\end{example}

\subsection{The weight distribution of $\mathcal{C}$ for  $ v_2(k)<v_2(m)$}
In this subsection, we always assume that $v_2(k)<v_2(m)$. To determine the value distribution of $S(\alpha,\beta)$, we  need some identities on $S(\alpha,\beta)$.
\begin{lemma}\label{lem:4gexingzhi}
With the notations given above, we have the following result:
\begin{enumerate}
  \item $\sum\limits_{\alpha,\beta\in \mathbb{F}_{p^m}}S(\alpha,\beta)=2p^{2m}.$
  \item $\sum\limits_{\alpha,\beta\in \mathbb{F}_{p^m}}S(\alpha,\beta)^2=\left\{
                                                            \begin{array}{ll}
                                                             4 p^{3m}, & \hbox{if $p^k \equiv 1\ ({\rm mod}\ 4)$;} \\
                                                              4 p^{2m}, & \hbox{if $p^k \equiv 3\ ({\rm mod}\ 4)$.}
                                                            \end{array}
                                                          \right.$
  \item $\sum\limits_{\alpha,\beta\in \mathbb{F}_{p^m}}S(\alpha,\beta)^3=\left\{
                                                             \begin{array}{ll}
                                                               2p^{2m}(7p^{m}-3)+2p^{2m+d}(p^m-1), & \hbox{if $p^k \equiv 1\ ({\rm mod}\ 4)$;} \\
2p^{2m}(p^m+3)+2p^{2m+d}(p^m-1), & \hbox{if $p^k \equiv 3\ ({\rm mod}\ 4)$.}
                                                             \end{array}
                                                           \right.$
  \item $(p^d-1)\sum\limits_{(\alpha,\beta)\in N_1}S(\alpha,\beta)+(p^{2d}-1)\sum\limits_{(\alpha,\beta)\in N_2}S(\alpha,\beta)=p^{m}(p^d-1)(p^m-1).$
\end{enumerate}
\end{lemma}
\pf $1$. We compute
\begin{eqnarray*}
  \sum_{\alpha,\beta\in \mathbb{F}_{p^m}}S(\alpha,\beta) &=& \sum_{\alpha,\beta\in\mathbb{F}_{p^m}}\left(T(\alpha,\beta)+ T(\pi^{\frac{p^k+1}{2}}\alpha,-\pi \beta)\right)\\
  &=&\sum_{\alpha,\beta\in\mathbb{F}_{p^m}}T(\alpha,\beta)+\sum_{\alpha',\beta'\in\mathbb{F}_{p^m}}T(\alpha',\beta')\\
&=& 2\sum_{x\in \mathbb{F}_{p^m}}\sum_{\alpha\in \mathbb{F}_{p^m}}\zeta_p^{ {\rm Tr}_1^m(\alpha x^{p^k+1})}\sum_{\beta\in \mathbb{F}_{p^m}}\zeta_p^{ {\rm Tr}_1^m(\beta x^2)} \\
   &=& 2p^{2m}.
\end{eqnarray*}

$2$. By (\ref{eq:Salphabata}), we get that
\begin{equation}\label{eq:S^2}
\sum_{\alpha,\beta\in \mathbb{F}_{p^m}}S(\alpha,\beta)^2
=\sum_{\alpha,\beta\in\mathbb{F}_{p^m}}\left(T(\alpha,\beta)^2+2T(\alpha,\beta)T(\pi^{\frac{p^k+1}{2}}\alpha,-\pi \beta)+ T(\pi^{\frac{p^k+1}{2}}\alpha,-\pi \beta)^2\right).
\end{equation}
On one hand,
\begin{eqnarray}\label{eq:Mp2m}
\nonumber \sum_{\alpha,\beta\in\mathbb{F}_{p^m}}T(\alpha,\beta)^2 &=& \sum_{x,y\in\mathbb{F}_{p^m}}\sum_{\alpha\in\mathbb{F}_{p^m}}\zeta_p^{ {\rm Tr}_1^m(\alpha (x^{p^k+1}+y^{p^k+1}))}\sum_{\beta\in\mathbb{F}_{p^m}}\zeta_p^{ {\rm Tr}_1^m(\alpha (x^{2}+y^{2}))} \\
 &=& Mp^{2m},
\end{eqnarray}
where $M=\#\{(x,y)\in \mathbb{F}_{p^m}^2\mid x^2+y^2=0, x^{p^k+1}+y^{p^k+1}=0\}$. Together with Lemma~\ref{lem:E1} ii) and  (\ref{eq:Mp2m}), we obtain that
\[\sum_{\alpha,\beta\in\mathbb{F}_{p^m}}T(\alpha,\beta)^2=\left\{
                                                            \begin{array}{ll}
                                                              p^{2m}(2p^m-1), & \hbox{if $p^k \equiv 1\ ({\rm mod}\ 4)$;} \\
                                                               p^{2m}, & \hbox{if $p^k \equiv 3\ ({\rm mod}\ 4)$.}
                                                            \end{array}
                                                          \right.
 \]
Similarly, \[\sum_{\alpha,\beta\in\mathbb{F}_{p^m}}T(\pi^{\frac{p^k+1}{2}}\alpha,-\pi \beta)^2=\left\{
                                                            \begin{array}{ll}
                                                              p^{2m}(2p^m-1), & \hbox{if $p^k \equiv 1\ ({\rm mod}\ 4)$;} \\
                                                               p^{2m}, & \hbox{if $p^k \equiv 3\ ({\rm mod}\ 4)$.}
                                                            \end{array}
                                                          \right.
 \]
On the other hand,
\begin{eqnarray*}
   && \sum_{\alpha,\beta\in\mathbb{F}_{p^m}}T(\alpha,\beta)T(\pi^{\frac{p^k+1}{2}}\alpha,-\pi \beta) \\
   &=& \sum_{x,y\in\mathbb{F}_{p^m}}\sum_{\alpha\in\mathbb{F}_{p^m}}\zeta_p^{ {\rm Tr}_1^m(\alpha( x^{p^k+1}+\pi^{\frac{p^k+1}{2}}y^{p^k+1}))}\sum_{\beta\in\mathbb{F}_{p^m}}\zeta_p^{ {\rm Tr}_1^m(\beta (x^{2}-\pi y^{2}))}  \\
   &=&  \#\{(x,y)\in \mathbb{F}_{p^m}^2\mid x^2-\pi y^2=0, x^{p^k+1}+\pi^{\frac{p^k+1}{2}}y^{p^k+1}=0\}\cdot p^{2m}\\
   &=&  \#\{(x,y)\in \mathbb{F}_{p^m}^2\mid x^2-\pi y^2=0, 2\pi^{\frac{p^k+1}{2}}y^{p^k+1}=0\}\cdot p^{2m}\\
&=& p^{2m}.
\end{eqnarray*}
Hence, from (\ref{eq:S^2}), we get
\[\sum_{\alpha,\beta\in \mathbb{F}_{p^m}}S(\alpha,\beta)^2 =\left\{
                                                            \begin{array}{ll}
                                                             4 p^{3m}, & \hbox{if $p^k \equiv 1\ ({\rm mod}\ 4)$;} \\
                                                              4 p^{2m}, & \hbox{if $p^k \equiv 3\ ({\rm mod}\ 4)$.}
                                                            \end{array}
                                                          \right.
 \]

$3$. By (\ref{eq:Salphabata}),
\begin{eqnarray}\label{eq:S^3}
\nonumber  \sum_{\alpha,\beta\in \mathbb{F}_{p^m}}S(\alpha,\beta)^3  &=& \sum_{\alpha,\beta\in\mathbb{F}_{p^m}}(T(\alpha,\beta)^3+3T(\alpha,\beta)^2T(\pi^{\frac{p^k+1}{2}}\alpha,-\pi\beta) \\
   & &+ 3T(\alpha,\beta)T(\pi^{\frac{p^k+1}{2}}\alpha,-\pi \beta)^2+T(\pi^{\frac{p^k+1}{2}}\alpha,-\pi \beta)^3).
\end{eqnarray}
It is easy to check that
\begin{equation}\label{eq:1}
\sum_{\alpha,\beta\in\mathbb{F}_{p^m}}T(\alpha,\beta)^3=\sum_{\alpha,\beta\in\mathbb{F}_{p^m}}
T(\pi^{\frac{p^k+1}{2}}\alpha,-\pi\beta)^3
\end{equation}
and
\begin{equation}\label{eq:2}
\sum_{\alpha,\beta\in\mathbb{F}_{p^m}}T(\alpha,\beta)^2T(\pi^{\frac{p^k+1}{2}}\alpha,-\pi\beta)=\sum_{\alpha,\beta\in\mathbb{F}_{p^m}}T(\alpha,\beta)T(\pi^{\frac{p^k+1}{2}}\alpha,-\pi \beta)^2.
\end{equation}
By Lemma~\ref{lem:zhishuhefenbu} ii),
\begin{equation}\label{eq:3}
\sum_{\alpha,\beta\in\mathbb{F}_{p^m}}T(\alpha,\beta)^3=p^{3m}+p^{2m+d}(p^m-1).
\end{equation}
By Lemma~\ref{lem:E2}, we have
\begin{eqnarray}\label{eq:4}
\nonumber   & &\sum_{\alpha,\beta\in\mathbb{F}_{p^m}}T(\alpha,\beta)^2T(\pi^{\frac{p^k+1}{2}}\alpha,-\pi\beta)  \\
\nonumber    &=& \sum_{x,y,z\in\mathbb{F}_{p^m}}\sum_{\alpha\in\mathbb{F}_{p^m}}\zeta_p^{ {\rm Tr}_1^m(\alpha( x^{p^k+1}+y^{p^k+1}+\pi^{\frac{p^k+1}{2}}z^{p^k+1}))}\sum_{\beta\in\mathbb{F}_{p^m}}\zeta_p^{ {\rm Tr}_1^m(\beta (x^{2}+y^2-\pi z^{2}))} \\
\nonumber    &=& \#\{(x,y,z)\in \mathbb{F}_{p^m}^3\mid x^2+y^2-\pi z^2=0, x^{p^k+1}+y^{p^k+1}+\pi^{\frac{p^k+1}{2}}z^{p^k+1})=0\}\cdot p^{2m} \\
   &=& \left\{
         \begin{array}{ll}
           (2p^m-1)p^{2m}, & \hbox{if $p^k \equiv 1\ ({\rm mod}\ 4)$;} \\
           p^{2m}, & \hbox{if $p^k \equiv 3\ ({\rm mod}\ 4)$.}
         \end{array}
       \right.
\end{eqnarray}
Combining (\ref{eq:S^3})-(\ref{eq:4}), we get
\[\sum_{\alpha,\beta\in \mathbb{F}_{p^m}}S(\alpha,\beta)^3=\left\{
                                                             \begin{array}{ll}
                                                               2p^{2m}(7p^{m}-3)+2p^{2m+d}(p^m-1), & \hbox{if $p^k \equiv 1\ ({\rm mod}\ 4)$;} \\
2p^{2m}(p^m+3)+2p^{2m+d}(p^m-1), & \hbox{if $p^k \equiv 3\ ({\rm mod}\ 4)$.}
                                                             \end{array}
                                                           \right.
\]

$4$. By  (\ref{eq:alphabetafenbu}) and (\ref{eq:Salphabata}),
\begin{eqnarray}\label{eq:11}
\nonumber   && (p^d-1)\sum\limits_{(\alpha,\beta)\in N_1}S(\alpha,\beta)+(p^{2d}-1)\sum\limits_{(\alpha,\beta)\in N_2}S(\alpha,\beta) \\
   &=& \sum_{x,\alpha \in \mathbb{F}_{p^m}^*}\left(T(\alpha,\psi(\alpha,x))+T(\pi^{\frac{p^k+1}{2}}\alpha,-\pi \psi(\alpha,x))\right).
\end{eqnarray}
By (\ref{eq:psix}),
\begin{eqnarray}\label{eq:12}
\nonumber    \sum_{x,\alpha \in \mathbb{F}_{p^m}^*}T(\alpha,\psi(\alpha,x))
  &=& \sum_{x,\alpha \in \mathbb{F}_{p^m}^*}\sum_{y\in \mathbb{F}_{p^m}}\zeta_p^{ {\rm Tr}_1^m(\alpha y^{p^k+1} -\frac{1}{2}x^{-1}(\alpha x^{p^k}+ \alpha^{p^{m-k}}x^{p^{m-k}})y^2)}\\
\nonumber   &=& \sum_{x,\alpha \in \mathbb{F}_{p^m}^*}\sum_{y\in \mathbb{F}_{p^m}}\zeta_p^{ {\rm Tr}_1^m(\alpha y^{p^k+1} -\frac{1}{2}x^{-1}\alpha x^{p^k}y^2-\frac{1}{2}x^{-p^k} \alpha x y^{2p^k})}\\
\nonumber&=& \sum_{x\in\mathbb{F}_{p^m}^*,y\in\mathbb{F}_{p^m}}\sum_{\alpha \in \mathbb{F}_{p^m}^* }\zeta_p^{ {\rm Tr}_1^m(-\frac{1}{2}\alpha (y^2x^{1-p^k}( x^{p^k-1}-y^{p^k-1})^2))}\\
\nonumber&=& (p^m-1)M-(p^m(p^m-1)-M)\\
&=&p^mM-p^{2m}+p^m,
\end{eqnarray}
where
\begin{eqnarray}\label{eq:13}
\nonumber  M&=&\#\{x\in \mathbb{F}_{p^m}^*,y\in \mathbb{F}_{p^m}\mid y^2( x^{p^k-1}-y^{p^k-1})^2=0\}\\
\nonumber   &=& \#\{x\in \mathbb{F}_{p^m}^*,y\in \mathbb{F}_{p^m}\mid  \ y=\lambda x, {\rm where} \ \lambda\in \mathbb{F}_{p^d}\} \\
   &=&  (p^m-1)p^d.
\end{eqnarray}
On the other hand, by (\ref{eq:alphabetafenbu}),
\begin{eqnarray}\label{eq:14}
\nonumber    \sum_{x,\alpha \in \mathbb{F}_{p^m}^*}T(\pi^{\frac{p^k+1}{2}}\alpha,-\pi \psi(\alpha,x))&=& \sum_{x,\alpha \in \mathbb{F}_{p^m}^*}\sum_{y\in \mathbb{F}_{p^m}}\zeta_p^{ {\rm Tr}_1^m(\pi^{\frac{p^k+1}{2}}\alpha y^{p^k+1} +\pi\frac{1}{2}x^{-1}(\alpha x^{p^k}+ \alpha^{p^{m-k}}x^{p^{m-k}})y^2)}\\
\nonumber&=& \sum_{x\in\mathbb{F}_{p^m}^*,y\in\mathbb{F}_{p^m}}\sum_{\alpha \in \mathbb{F}_{p^m}^* }\zeta_p^{ {\rm Tr}_1^m(\frac{1}{2}\pi \alpha (y^2x^{1-p^k}( x^{p^k-1}+\pi^{\frac{p^k-1}{2}}y^{p^k-1})^2))}\\
\nonumber&=& (p^m-1)I-(p^m(p^m-1)-I)\\
&=&p^mI-p^{2m}+p^m,
\end{eqnarray}
where
\begin{eqnarray*}
  I&=&\#\{x\in \mathbb{F}_{p^m}^*,y\in \mathbb{F}_{p^m}\mid y^2( x^{p^k-1}+\pi^{\frac{p^k-1}{2}}y^{p^k-1})^2=0\}\\
  &=& \#\{x\in \mathbb{F}_{p^m}^*,y\in \mathbb{F}_{p^m}\mid y=0 \ {\rm or} \ (\frac{x}{y})^{p^k-1}=-\pi^{\frac{p^k-1}{2}}\}.
\end{eqnarray*}
Since $\frac{k}{d}$ is odd and $m$ is even, then $p^d-1 \nmid \frac{p^k-1}{2}+\frac{p^m-1}{2}$, which implies that there is no solution of $(\frac{x}{y})^{p^k-1}=-\pi^{\frac{p^k-1}{2}}$. Hence, $I=p^m-1$. Together with (\ref{eq:11}), (\ref{eq:12}) and (\ref{eq:14}), we have that \[ (p^d-1)\sum\limits_{(\alpha,\beta)\in N_1}S(\alpha,\beta)+(p^{2d}-1)\sum\limits_{(\alpha,\beta)\in N_2}S(\alpha,\beta)=p^{m}(p^d-1)(p^m-1).\]\EOP

\begin{theorem}\label{th:Svaluedistribution1}
With the notations given above. Then the value distribution of $ S(\alpha,\beta)$ defined by (\ref{eq:Salphabata}) is as follows:
\begin{center}
\begin{tabular}{|l|l|}
  \hline
  value & frequency \\ \hline
  $2p^m$ & $1$ \\ \hline
  $0$ & $\frac{1}{2}(p^{m+d}-3p^m+p^d+1)(p^m-1)/(p^d-1)$ \\ \hline
  $\pm (p^d-1)p^{\frac{m}{2}}$ & $(p^{\frac{m}{2}-d}\pm1)(p^\frac{m}{2}\mp1)(p^m-1)/(p^{2d}-1)$ \\ \hline
  $ (\pm\sqrt{q^*} -1)p^{\frac{m}{2}}$ & $\frac{1}{2}p^{\frac{m}{2}-d}(p^{\frac{m}{2}}-1)(p^m-1)$ \\ \hline
  $ (\pm\sqrt{q^*} +1)p^{\frac{m}{2}}$ & $\frac{1}{2}p^{\frac{m}{2}-d}(p^{\frac{m}{2}}+1)(p^m-1)$ \\ \hline
  $\pm2p^{\frac{m}{2}}$  & $\frac{1}{4}(p^{\frac{m}{2}}\pm1)^2(p^d-1)(p^{m}-1)/(p^d+1)$ \\
  \hline
\end{tabular}
\end{center}
\end{theorem}
\pf From Lemma~\ref{lem:rfrg} and Lemma~\ref{lem:zhishuhefenbu}, we get that
\begin{equation}\label{eq:th21}
    \left\{
      \begin{array}{ll}
        n_{1,0}=n_{1,1,0}+n_{-1,1,0}+n_{1,1,1}+n_{-1,1,1}+n_{1,1,2}+n_{-1,1,2},  \\
        n_{-1,0}=n_{1,-1,0}+n_{-1,-1,0}+n_{1,-1,1}+n_{-1,-1,1}+n_{1,-1,2}+n_{-1,-1,2}.
      \end{array}
    \right.
\end{equation}
By (\ref{eq:Salphabata}) and Lemma~\ref{lem:rfrg},
\begin{equation}\label{eq:th22}
    \left\{
      \begin{array}{ll}
        n_{1,1}=n_{1,1,1}+n_{1,-1,1},  \\
        n_{-1,1}=n_{-1,1,1}+n_{-1,-1,1},\\
        n_{1,2}=n_{1,1,2}+n_{1,-1,2},  \\
        n_{-1,2}=n_{-1,1,2}+n_{-1,-1,2}.
      \end{array}
    \right.
\end{equation}
Note that if the rank of ${\rm Tr}_d^m(\alpha x^{p^k+1}+\beta x^2)$  is odd and we compute the value of $T(\alpha,\beta)$ and $T(a\alpha,a\beta)$, where $a$ is a nonsquare element in $\mathbb{F}_q$. Then, by Lemma~\ref{lem:erciquzhi}, we have
\[T(a\alpha,a\beta)=-T(\alpha,\beta), \]
and since $s$ is even , we get
\begin{equation}\label{eq:th23}
    \left\{
      \begin{array}{ll}
        n_{1,1,1}=n_{-1,1,1},  \\
        n_{1,-1,1}=n_{-1,-1,1}.
      \end{array}
    \right.
\end{equation}
On the other hand, we have
\begin{equation}\label{eq:th24}
  \begin{array}{ll}
    &\sum\limits_{\alpha,\beta \in \mathbb{F}_{p^m}}S(\alpha,\beta) =2p^m+2p^{\frac{m}{2}}( n_{1,1,0}-n_{-1,-1,0}+2(\sqrt{q^*}+1)(n_{1,1,1}-n_{-1,-1,1})+\\  &2(\sqrt{q^*}-1)(n_{1,-1,1}-n_{-1,1,1})+(p^d+1)(n_{1,1,2}-n_{-1,-1,2})+(p^d-1)(n_{1,-1,2}-n_{-1,1,2})),
  \end{array}
\end{equation}
\begin{equation}\label{eq:th25}
  \begin{array}{ll}
    &\sum\limits_{\alpha,\beta \in \mathbb{F}_{p^m}}S(\alpha,\beta)^2=4p^{2m}+2p^m (2(n_{1,1,0}+n_{-1,-1,0})+(\sqrt{q^*}+1)^2(n_{1,1,1}+n_{-1,-1,1})+\\
&(\sqrt{q^*}-1)^2(n_{1,-1,1}-n_{-1,1,1})+(p^d+1)^2(n_{1,1,2}+n_{-1,-1,2})+ (p^d-1)^2(n_{1,-1,2}+n_{-1,1,2})),
  \end{array}
\end{equation}
\begin{equation}\label{eq:th26}
  \begin{array}{ll}
    & \sum\limits_{\alpha,\beta \in \mathbb{F}_{p^m}}S(\alpha,\beta)^3=8p^{3m}+2p^{\frac{3m}{2}}(4( n_{1,1,0}-n_{-1,-1,0})+(\sqrt{q^*}+1)^3(n_{1,1,1}-n_{-1,-1,1})+\\
&(\sqrt{q^*}-1)^3(n_{1,-1,1}-n_{-1,1,1})+(p^d+1)^3(n_{1,1,2}-n_{-1,-1,2})+(p^d-1)^3(n_{1,-1,2}-n_{-1,1,2}))   ,
  \end{array}
\end{equation}
\begin{equation}\label{eq:th27}
  \begin{array}{ll}
&(p^d-1)\sum\limits_{(\alpha,\beta)\in N_1}S(\alpha,\beta)+(p^{2d}-1)\sum\limits_{(\alpha,\beta)\in N_2}S(\alpha,\beta)=(p^d-1)p^{\frac{m}{2}}((\sqrt{q^*}+1)(n_{1,1,1}-n_{-1,-1,1})+\\
&(\sqrt{q^*}-1)(n_{1,-1,1}-n_{-1,1,1})+(p^d+1)^2(n_{1,1,2}-n_{-1,-1,2})+(p^{2d}-1)(n_{1,-1,2}-n_{-1,1,2})).
  \end{array}
\end{equation}
Applying Lemma~\ref{lem:4gexingzhi} $1)-4)$ and solving the system equations
consisting of (\ref{eq:th21})-(\ref{eq:th27}), we get
\[n_{1,1,0}=\frac{1}{4}(p^{\frac{m}{2}}+1)^2(p^d-1)(p^{m}-1)/(p^d+1),\]
\[n_{-1,-1,0}=\frac{1}{4}(p^{\frac{m}{2}}-1)^2(p^d-1)(p^{m}-1)/(p^d+1),\]
\[n_{1,-1,0}=n_{-1,1,0}=\frac{1}{4}(p^{m+d}-3p^m+p^d+1)(p^m-1)/(p^d-1),\]
\[n_{1,1,1}=n_{-1,1,1}=\frac{1}{4}p^{\frac{m}{2}-d}(p^{\frac{m}{2}}+1)(p^m-1),\]
\[n_{1,-1,1}=n_{-1,-1,1}=\frac{1}{4}p^{\frac{m}{2}-d}(p^{\frac{m}{2}}-1)(p^m-1),\]
\[n_{1,-1,2}=\frac{1}{2}(p^{\frac{m}{2}-d}+1)(p^\frac{m}{2}-1)(p^m-1)/(p^{2d}-1),\]
\[n_{-1,1,2}=\frac{1}{2}(p^{\frac{m}{2}-d}-1)(p^\frac{m}{2}+1)(p^m-1)/(p^{2d}-1),\]
\[n_{1,1,2}=n_{-1,-1,2}=0. \]
This finishes the proof. \EOP

Therefore, we can determine the weight distribution of cyclic code $\mathcal{C}$.
\begin{theorem}\label{th:Cdistribution2}
With the notations given above.

i)  If $k$ is odd, then $\mathcal{C}$ defined by (\ref{eq:cycliccode}) is a cyclic code over $\mathbb{F}_p$ with length $p^m-1$ and dimension $2m$. Moreover, the weight distribution of cyclic code $ \mathcal{C}$  is given in Table~$2$.

\begin{table}[!h]
\tabcolsep 0pt
\caption{ weight distribution of $ \mathcal{C}$ for  odd $k$}
\vspace*{-12pt}
\begin{center}
\def\temptablewidth{0.9\textwidth}
{\rule{\temptablewidth}{1pt}}
\begin{tabular*}{\temptablewidth}{@{\extracolsep{\fill}}cc}
value & frequency \\ \hline
  $0$ & $1$ \\
  $p^m-p^{m-1}$ & $\frac{1}{2}(p^{m+d}-3p^m+p^d+1)(p^m-1)/(p^d-1)$ \\
  $p^{m-1}(p-1)- \frac{1}{2}(p-1)(p^d-1)p^{\frac{m}{2}-1}$ & $(p^{\frac{m}{2}-d}+1)(p^\frac{m}{2}-1)(p^m-1)/(p^{2d}-1)$ \\
$p^{m-1}(p-1)+ \frac{1}{2}(p-1)(p^d-1)p^{\frac{m}{2}-1}$ & $(p^{\frac{m}{2}-d}-1)(p^\frac{m}{2}+1)(p^m-1)/(p^{2d}-1)$ \\
  $p^{m-1}(p-1)+\frac{1}{2}(p-1)p^{\frac{m}{2}-1}$ & $p^{\frac{m}{2}-d}(p^{\frac{m}{2}}-1)(p^m-1)$ \\
  $p^{m-1}(p-1)- \frac{1}{2}(p-1)p^{\frac{m}{2}-1}$ & $p^{\frac{m}{2}-d}(p^{\frac{m}{2}}+1)(p^m-1)$ \\
  $p^{m-1}(p-1)+(p-1)p^{\frac{m}{2}-1}$  & $\frac{1}{4}(p^{\frac{m}{2}}-1)^2(p^d-1)(p^{m}-1)/(p^d+1)$ \\
  $p^{m-1}(p-1)-(p-1)p^{\frac{m}{2}-1}$  & $\frac{1}{4}(p^{\frac{m}{2}}+1)^2(p^d-1)(p^{m}-1)/(p^d+1)$ \\
\end{tabular*}
{\rule{\temptablewidth}{1pt}}
\end{center}
\end{table}

ii) If  $k$ is even, then $\mathcal{C}$ defined by (\ref{eq:cycliccode}) is a cyclic code over $\mathbb{F}_p$ with length $p^m-1$ and dimension $2m$. Moreover, the weight distribution of cyclic code $ \mathcal{C}$  is given in Table~$3$.

\begin{table}[!h]
\tabcolsep 0pt
\caption{ weight distribution of $ \mathcal{C}$ for  even $k$}
\vspace*{-12pt}
\begin{center}
\def\temptablewidth{0.9\textwidth}
{\rule{\temptablewidth}{1pt}}
\begin{tabular*}{\temptablewidth}{@{\extracolsep{\fill}}cc}
Weight & Frequency \\   \hline
 $0$ & $1$ \\
  $p^m-p^{m-1}$ & $\frac{1}{2}(p^{m+d}-3p^m+p^d+1)(p^m-1)/(p^d-1)$ \\
  $p^{m-1}(p-1)+ \frac{1}{2}(p-1)(p^d-1)p^{\frac{m}{2}-1}$ & $(p^{\frac{m}{2}-d}-1)(p^\frac{m}{2}+1)(p^m-1)/(p^{2d}-1)$ \\
$p^{m-1}(p-1)- \frac{1}{2}(p-1)(p^d-1)p^{\frac{m}{2}-1}$ & $(p^{\frac{m}{2}-d}+1)(p^\frac{m}{2}-1)(p^m-1)/(p^{2d}-1)$ \\
  $p^{m-1}(p-1)- \frac{1}{2}(p-1)(p^{\frac{d}{2}} -1)p^{\frac{m}{2}-1}$ & $\frac{1}{2}p^{\frac{m}{2}-d}(p^{\frac{m}{2}}-1)(p^m-1)$ \\
$p^{m-1}(p-1)+ \frac{1}{2}(p-1)(p^{\frac{d}{2}}+1)p^{\frac{m}{2}-1}$ & $\frac{1}{2}p^{\frac{m}{2}-d}(p^{\frac{m}{2}}-1)(p^m-1)$ \\
  $p^{m-1}(p-1)- \frac{1}{2}(p-1)(p^{\frac{d}{2}} +1)p^{\frac{m}{2}-1}$ & $\frac{1}{2}p^{\frac{m}{2}-d}(p^{\frac{m}{2}}+1)(p^m-1)$ \\
 $p^{m-1}(p-1)+ \frac{1}{2}(p-1)(p^{\frac{d}{2}} -1)p^{\frac{m}{2}-1}$ & $\frac{1}{2}p^{\frac{m}{2}-d}(p^{\frac{m}{2}}+1)(p^m-1)$ \\
  $p^{m-1}(p-1)+(p-1)p^{\frac{m}{2}-1}$  & $\frac{1}{4}(p^{\frac{m}{2}}-1)^2(p^d-1)(p^{m}-1)/(p^d+1)$ \\
  $p^{m-1}(p-1)-(p-1)p^{\frac{m}{2}-1}$  & $\frac{1}{4}(p^{\frac{m}{2}}+1)^2(p^d-1)(p^{m}-1)/(p^d+1)$ \\
\end{tabular*}
{\rule{\temptablewidth}{1pt}}
\end{center}
\end{table}

\end{theorem}
\pf By (\ref{eq:wtcab}), we have
\[wt(\textbf{c}(\alpha,\beta))=p^m-p^{m-1}-\frac{1}{2p}\sum_{u\in \mathbb{F}_p^*}S(u\alpha,u\beta).\]

If $k$ is odd, then $d$ is odd and $u_p=\pi^{\frac{p^m-1}{p-1}}=\pi^{\frac{p^m-1}{p^d-1}\cdot \frac{p^d-1}{p-1}}$ is a nonsquare element in $\mathbb{F}_{p^d}$. So we have $\eta_d(u_p)=-1$. If $u\in SQ_p$, then $u^{\frac{p^k+1}{2}}=u$. Hence, by Lemma~\ref{lem:erciquzhi}, we have
\begin{eqnarray*}
 && \sum_{u\in \mathbb{F}_p^*}S(u\alpha,u\beta)\\
   &=&  \sum_{u\in \mathbb{F}_p^*}\sum_{x\in \mathbb{F}_{p^m}}\left(\zeta_p^{ {\rm Tr}_1^m(u\alpha x^{p^k+1}+u\beta x^2)}+\zeta_p^{ {\rm Tr}_1^m(u\alpha\pi^{\frac{p^k+1}{2}} x^{p^k+1}-u\beta\pi x^2)}\right)\\
&=& \sum_{u\in SQ_p}\sum_{x\in \mathbb{F}_{p^m}}\left(\zeta_p^{ {\rm Tr}_1^m(\alpha (u^{\frac{1}{2}}x)^{p^k+1}+\beta (u^{\frac{1}{2}}x)^2)}+\zeta_p^{ {\rm Tr}_1^m(\alpha\pi^{\frac{p^k+1}{2}} (u^{\frac{1}{2}}x)^{p^k+1}-\beta\pi (u^{\frac{1}{2}}x)^2)}\right)\\
& +&\sum_{u\in SQ_p}\sum_{x\in \mathbb{F}_{p^m}}\left(\zeta_p^{ u_p{\rm Tr}_1^m(\alpha (u^{\frac{1}{2}}x)^{p^k+1}+\beta (u^{\frac{1}{2}}x)^2)}+\zeta_p^{u_p {\rm Tr}_1^m(\alpha\pi^{\frac{p^k+1}{2}} (u^{\frac{1}{2}}x)^{p^k+1}-\beta\pi (u^{\frac{1}{2}}x)^2)}\right)\\
&=& \sum_{u\in SQ_p}\sum_{x\in \mathbb{F}_{p^m}}\left(\zeta_p^{ {\rm Tr}_1^m(\alpha x^{p^k+1}+\beta x^2)}+\zeta_p^{ {\rm Tr}_1^m(\alpha\pi^{\frac{p^k+1}{2}} x^{p^k+1}-\beta\pi x^2)}\right)\\
& +&\sum_{u\in SQ_p}\sum_{x\in \mathbb{F}_{p^m}}\left(\zeta_p^{ u_p{\rm Tr}_1^m(\alpha x^{p^k+1}+\beta x^2)}+\zeta_p^{u_p {\rm Tr}_1^m(\alpha\pi^{\frac{p^k+1}{2}} x^{p^k+1}-\beta\pi x^2)}\right)\\
&=& \frac{p-1}{2}\left( (1+\eta_d(u_p)^r\sum_{x\in \mathbb{F}_{p^m}}\zeta_p^{ {\rm Tr}_1^m(\alpha x^{p^k+1}+\beta x^2)}+(1+\eta_d(u_p)^{r'}\sum_{x\in \mathbb{F}_{p^m}}\zeta_p^{ {\rm Tr}_1^m(\alpha\pi^{\frac{p^k+1}{2}} x^{p^k+1}-\beta\pi x^2)}\right),\\
\end{eqnarray*}
where $r$ and $r'$ are the rank of ${\rm Tr}_d^m(\alpha x^{p^k+1}+\beta x^2)$ and ${\rm Tr}_d^m(\alpha\pi^{\frac{p^k+1}{2}} x^{p^k+1}-\beta\pi x^2)$, respectively. Note that $\eta_d(u_p)=-1$ and $s$ is even.  By Lemma~\ref{lem:rfrg}, we have if $r=s-1$, then $r'=s$ and $$\sum_{u\in \mathbb{F}_p^*}S(u\alpha,u\beta)=(p-1)\sum_{x\in \mathbb{F}_{p^m}}\zeta_p^{ {\rm Tr}_1^m(\alpha\pi^{\frac{p^k+1}{2}} x^{p^k+1}-\beta\pi x^2)}.$$
If $r'=s-1$, then $r=s$ and $$\sum_{u\in \mathbb{F}_p^*}S(u\alpha,u\beta)=(p-1)\sum_{x\in \mathbb{F}_{p^m}}\zeta_p^{ {\rm Tr}_1^m(\alpha x^{p^k+1}+\beta x^2)}.$$
Otherwise, $$\sum_{u\in \mathbb{F}_p^*}S(u\alpha,u\beta)=(p-1)\sum_{x\in \mathbb{F}_{p^m}}\left(\zeta_p^{ {\rm Tr}_1^m(\alpha x^{p^k+1}+\beta x^2)}+\zeta_p^{ {\rm Tr}_1^m(\alpha\pi^{\frac{p^k+1}{2}} x^{p^k+1}-\beta\pi x^2)}\right).$$
Therefore, by Theorem~\ref{th:Svaluedistribution1}, we obtain the weight distribution of cyclic code $\mathcal{C}$ for odd $k$.

If $k$ is even, then $u^{\frac{p^k+1}{2}}=u$ for all $u\in \mathbb{F}_p$. Since $m$ is even, then $u$ is a square element in $\mathbb{F}_{p^m}$. Hence, we have
\begin{eqnarray*}
  S(u\alpha,u\beta)
   &=&  \sum_{x\in \mathbb{F}_{p^m}}\left(\zeta_p^{ {\rm Tr}_1^m(u\alpha x^{p^k+1}+u\beta x^2)}+\zeta_p^{ {\rm Tr}_1^m(u\alpha\pi^{\frac{p^k+1}{2}} x^{p^k+1}-u\beta\pi x^2)}\right)\\
&=& \sum_{x\in \mathbb{F}_{p^m}}\left(\zeta_p^{ {\rm Tr}_1^m(\alpha (u^{\frac{1}{2}}x)^{p^k+1}+\beta (u^{\frac{1}{2}}x)^2)}+\zeta_p^{ {\rm Tr}_1^m(\alpha\pi^{\frac{p^k+1}{2}} (u^{\frac{1}{2}}x)^{p^k+1}-\beta\pi (u^{\frac{1}{2}}x)^2)}\right)\\
&=&  \sum_{x\in \mathbb{F}_{p^m}}\left(\zeta_p^{ {\rm Tr}_1^m(\alpha x^{p^k+1}+\beta x^2)}+\zeta_p^{ {\rm Tr}_1^m(\alpha\pi^{\frac{p^k+1}{2}} x^{p^k+1}-\beta\pi x^2)}\right)=S(\alpha,\beta).
\end{eqnarray*}
Therefore,
\[wt(\textbf{c}(\alpha,\beta))=p^m-p^{m-1}-\frac{p-1}{2p}S(\alpha,\beta).\]
By Theorem~\ref{th:Svaluedistribution1}, we get the weight distribution of cyclic code $\mathcal{C}$ for  even $k$.\EOP

In the following, we give an  example to verify the result in Theorem~\ref{th:Cdistribution2} for the case of odd $k$. For the case of even $k$, we are not able to give an example to verify the result in Theorem~\ref{th:Cdistribution2} because of our limited computation ability.
\begin{example}
Let $p=3$, $m=6$, $k=1$, the  code $\mathcal{C}$ is a $[728,12,468]$ cyclic
code over $\mathbb{F}_3$ with weight enumerator
\[1+95004X^{468}+183456X^{477}+728X^{486}+170352X^{495}+81900X^{504},\]
which confirms the weight distribution in Table $2$.
\end{example}



\end{document}